\documentclass[12pt]{article}
\usepackage{amsmath,amssymb,a4wide,epsfig}
\input{epsf}
\newcommand\as{\alpha_{\mathrm{S}}}

\def\beq{\begin{equation}}
\def\eeq{\end{equation}}
\def\beeq{\begin{eqnarray}}
\def\eeeq{\end{eqnarray}}

\def\to{\rightarrow}

\def\ptZZ{p_{T}^{ZZ}}
\def\ZZ{ZZ}
\def\ptcut{p_{T{\rm cut}}}

\begin{document} 

\begin{titlepage}
\renewcommand{\thefootnote}{\fnsymbol{footnote}}
\begin{flushright}
CP3-08-01
\end{flushright}
\vspace*{1cm}
\begin{center}
{\Large \bf Higher-order QCD effects\\
\vskip .3cm in the Higgs to $ZZ$ search channel at the LHC}
\end{center}
\par \vspace{2mm}
\begin{center}
{\bf Rikkert Frederix}${}^{(a)}$ and {\bf Massimiliano Grazzini}${}^{(b)}$\\

\vspace{5mm}

${}^{(a)}$Center for Particle Physics and Phenomenology (CP3),\\
Universit\'e catholique de Louvain, B-1348 Louvain-la-Neuve, Belgium

\vspace{5mm}

${}^{(b)}$INFN, Sezione di Firenze\\ I-50019 Sesto Fiorentino,
Florence, Italy\\

\vspace{5mm}

\end{center}

\par \vspace{2mm}
\begin{center}
{\large \bf Abstract}
\end{center}
\begin{quote}
\pretolerance 10000
We present a consistent analysis of the signal
as well as the irreducible background for the search of the SM Higgs boson in the $ZZ$ decay channel at the LHC.
Soft-gluons effects are resummed up to
next-to-leading logarithmic accuracy,
and the results are compared to those obtained with fixed
order calculations and the MC@NLO event generator.
The soft-gluon effects are typically
modest but should be taken into account
when precise predictions are demanded. 
Our results show that
the signal over background ratio can be
significantly enhanced with a cut on the transverse momentum $\ptZZ$
of the $ZZ$ pair. We also introduce
a fully transverse angular variable that
could give information about the CP nature of the Higgs boson.
\end{quote}

\vspace*{\fill}
\begin{flushleft}
January 2008

\end{flushleft}
\end{titlepage}

\setcounter{footnote}{1}
\renewcommand{\thefootnote}{\fnsymbol{footnote}}

\section{Introduction}
\label{sec:intro}

The elucidation of the mechanism of electroweak symmetry breaking is
one of the main goals of the LHC physics program.  In the Standard
Model (SM) and several popular extensions such as SUSY, mass
generation is triggered by the Higgs mechanism, which predicts the
existence of (at least) one scalar state, the Higgs boson.  The search
for the Higgs at collider experiments has now being on-going for two
decades. The present direct lower limit of the Higgs mass in the SM is
114.4 GeV (at 95\% CL) \cite{Barate:2003sz}, while precision measurements point to a rather
light Higgs, $m_h \lesssim 200$ GeV \cite{Alcaraz:2007ri}.

At the LHC, the main production mechanism will be $gg \to H$, and if
$m_h>180$ GeV, the Higgs decay into two
Z bosons, $h\to ZZ$,
will provide one of the cleanest signatures at hadron colliders, {\em i.e.},
four leptons. Such a final state will allow a very accurate mass
reconstruction and the best of all possible discovery modes, a sharp
peak over a rather flat background. At this stage, accurate predictions
from theory will be helpful to design the best analysis but are not
essential to claim a discovery as data alone will provide all the
necessary information. However, to answer the key questions on the
nature of the discovered particle, such as its spin, CP nature and
couplings, accurate predictions for both signal and backgrounds will
be required.

As far as the Higgs signal is concerned, QCD corrections at the
next-to-leading order (NLO) have been known for some time
\cite{Dawson:1991zj,Spira:1995rr}: their effect increases the LO cross
section by about 80--100\%.  In recent years, even
next-to-next-to-leading order (NNLO) corrections have been computed,
first for the total cross section \cite{NNLOtotal}, and more recently
implemented in fully exclusive calculations
\cite{Anastasiou:2004xq,Catani:2007vq}. Note, however, that all the
NNLO results use the large-$m_{top}$ approximation, $m_{top}$ being the mass
of the top quark.

As far as $ZZ$ production is concerned, NLO corrections have been
known for some time
\cite{Ohnemus:1990za,Mele:1990bq,Ohnemus:1994ff}. More recent NLO
calculations exist that, using the one-loop helicity amplitudes of
Ref.~\cite{Dixon:1998py}, fully take into account spin correlations in
the $Z$ boson decay \cite{Dixon:1999di,Campbell:1999ah}.

The fixed-order calculations provide a reliable estimate of signal and
background cross sections and distributions as long as the scales
involved in the process are all of the same order. When the total
transverse momentum of the $ZZ$ pair is much smaller than its
invariant mass the validity of the fixed-order expansion may be
spoiled since the coefficients of the perturbative expansion can be
enhanced by powers of the large logarithmic terms, $\ln^n
M_{ZZ}/\ptZZ$.
In the case of the Higgs signal, the resummation of such contributions
has been performed up to next-to-next-to-leading-logarithmic (NNLL)
accuracy \cite{Bozzi:2003jy,Bozzi:2005wk,Bozzi:2007pn}.

The purpose of the present paper is twofold. We first consider
transverse momentum resummation for $ZZ$ production at the LHC.  The
resummation of such logarithmic contributions was first considered in
Ref.~\cite{Balazs:1998bm}.  Here we use the resummation formalism of
Refs.~\cite{Bozzi:2003jy,Bozzi:2005wk} together with the helicity
amplitudes of Ref.~\cite{Dixon:1998py} (including finite width effects
from the Z bosons, but neglecting single-resonant contributions).
Contrary to Ref.~\cite{Balazs:1998bm} we fully include the decay of
the $Z$ bosons,
keeping track of their polarization in the leptonic decay.
In the large $\ptZZ$ region we use LO perturbation theory ($\ZZ$+1
parton); in the region $\ptZZ\ll M_{ZZ}$ the large logarithmic
contributions are resummed to NLL accuracy. The present study
parallels the one performed in Ref.~\cite{Grazzini:2005vw} in the case
of $WW$ production.  By using these results, we perform a detailed
comparison of signal and background cross sections and distributions.

The paper is organized as follows. In Sect.~\ref{sec:zz} we analyze
the impact of transverse momentum resummation for $ZZ$ production. In
Sect.~\ref{sec:sb} we compare signal and background cross sections and
distributions for the search of a Higgs boson of mass $m_h=200$ GeV.
In Sect.~\ref{summa} we conclude with a summary of our results.

\section{Transverse-momentum resummation for $ZZ$ production}
\label{sec:zz}

In this Section we discuss the effect of transverse-momentum
resummation for $ZZ$ production at the LHC, and present a comparison
to fixed order NLO results obtained with MCFM \cite{Campbell:1999ah}
and to results obtained with MC@NLO \cite{MCatNLO}.

We consider the process $pp\to ZZ+X\to e^+e^-\mu^+\mu^-+X$ and perform
the all-order resummation of the logarithmically enhanced
contributions at small $\ptZZ$.  The implementation is completely
analogous to the case of $WW$ pair production discussed in
Ref.~\cite{Grazzini:2005vw} and is based on the formalism of
Refs.~\cite{Bozzi:2003jy,Bozzi:2005wk}. We refer the reader to the
above papers for the technical details.  The large logarithmic
contributions at small transverse momenta of the $ZZ$ pair are
resummed up to NLL accuracy. The result is then matched to the fixed
order LO calculation valid at large $\ptZZ$, to achieve NLL+LO
accuracy.

We recall that the formalism of Refs.~\cite{Bozzi:2003jy,Bozzi:2005wk}
enforces a unitarity constraint such that resummation effects vanish
when total cross sections are considered.  As a consequence, at NLL+LO
accuracy the integral of our resummed spectra coincides with the total
NLO cross section if no cuts are applied.

To compute the $ZZ$ cross section we use MRST2002 NLO parton densities
\cite{Martin:2002aw} and $\as$ evaluated at two-loop order.  Our
resummed predictions depend on renormalization, factorization and
resummation scales.
The resummation scale parametrizes the arbitrariness in the resummation procedure, and is set equal to the invariant mass $M_{ZZ}$
of the $ZZ$ pair. Variations around this central value can give an idea
of the size of yet uncalculated higher-order logarithmic contributions.
Renormalization and factorization scales are set to
$2M_Z$.  The latter choice allows us to exploit our unitarity
constraint and to exactly recover the total NLO cross section when no
cuts are applied.  At NLO we consistently use $\mu_F=\mu_R=2M_Z$ as
default choice, whereas in MC@NLO $\mu_F$ and $\mu_R$ are set to the
default choice, the average transverse mass of the $Z$ bosons.

The predictions of resummation are implemented in a partonic Monte
Carlo program which generates the full 5-body final state
($e^+e^-\mu^+\mu^-$ + 1 parton).  Nonetheless, since the resummed
cross section we use is inclusive over rapidity, we are not able to
apply the usual rapidity cuts on the leptons.  To the purpose of the
present work, we do not expect this limitation to be essential.

We start by considering the inclusive cross sections.
Our NLL+LO
result is 33.76 fb, and agrees with the NLO one (33.99 fb) to about
$1\%$.  With MC@NLO we obtain 34.60 fb. As expected, the MC@NLO cross
section is slightly larger because $ZZ$ production is calculated
in the narrow width approximation, while in the NLO and NLL+LO calculations, finite width effects are included.

\begin{figure}[tb]
  \begin{center}
    \epsfig{file=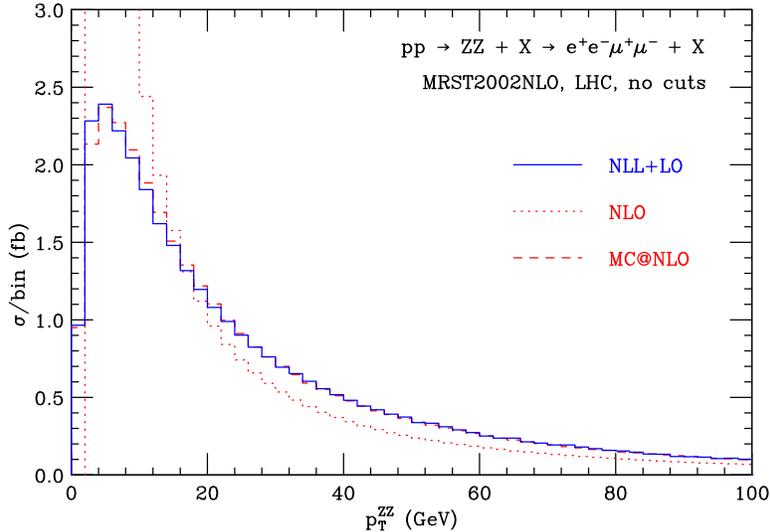, scale=0.65}
  \end{center}
  \vspace{-20pt}
  \caption{\label{fig:ptzz}{\em Comparison of the transverse momentum
      spectra of the $ZZ$ pair obtained at NLL+LO (solid) with MC@NLO
      (dashes) and NLO results (dots). No cuts are applied.}}
\end{figure}

In Fig.~\ref{fig:ptzz} we show
the corresponding $p^{ZZ}_T$ distribution,
computed at NLL+LO (solid), with MC@NLO (dashed) and at NLO (dots).
As is well known, the NLO result diverges to $+\infty$ as $\ptZZ\to 0$,
and this divergence is cancelled by the (negative) weight of the first bin, due to the virtual contribution.
On the contrary, the
NLL+LO and MC@NLO results are well behaved as $\ptZZ\to 0$
and are very close to each other, showing a peak
around $\ptZZ\sim 5$ GeV.

In order to study the perturbative uncertainties affecting our resummed
calculation, we have varied the renormalization and factorization scale by a factor 2 around the
central value. We find that the effect of $\mu_R$ and $\mu_F$ variations is rather small,
of the order of $\pm 1\%$, and comparable with the estimated accuracy of our
numerical code. Similar effects are found at NLO.

\begin{figure}[tb]
  \begin{center}
    \epsfig{file=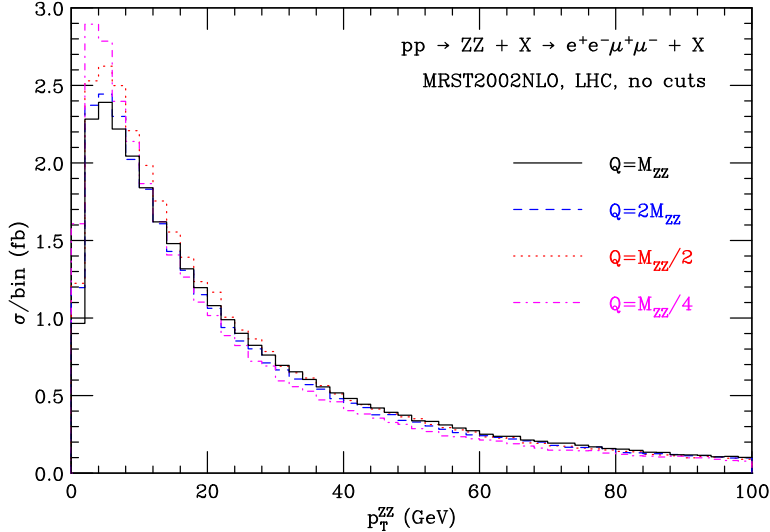, scale=0.65}
  \end{center}
  \vspace{-20pt}
  \caption{\label{fig:ptzz_Q}{\em Comparison of the transverse momentum
      spectra of the $ZZ$ pair obtained at NLL+LO for different values of the resummation scale $Q$. No cuts are applied.}}
\end{figure}

The dependence of our NLL+LO results on the resummation scale $Q$ is instead
stronger. In Fig.~\ref{fig:ptzz_Q} we show the NLL+LO prediction for different
choices of the resummation scale $Q$. We see that varying the resummation scale
the effect on the $\ptZZ$ spectrum is visible and amounts to about $\pm 10\%$
at the peak. For lower (higher) values of $Q$ the effect of resummation
is confined to smaller (larger) values of $\ptZZ$. Thanks to our unitarity constraint,
the total rate is instead insensitive to resummation scale variations,
within the numerical accuracy of our code.

As in the case of Higgs \cite{Bozzi:2005wk} and $WW$ \cite{Grazzini:2005vw} production, we find that the choice $Q=2M_{ZZ}$ gives (slightly) negative cross sections at very large $\ptZZ$. In order to
define a range of variations of $Q$, we prefer to avoid values that give a
bad behaviour at large $\ptZZ$. For this reason, in the following,
we will consider resummation scale variations in the range $M_{ZZ}/4\leq Q\leq M_{ZZ}$.

We now consider the selection cuts designed for the search of a Higgs
boson of mass $m_h=200$ GeV in the $e^+e^-\mu^+\mu^-$ channel
\cite{cmsnote}. The final-state leptons, ordered according to
decreasing $p_T$, should fulfil the following thresholds:
\begin{equation}
p_{T1}>22~{\rm GeV}~~~p_{T2}>20~{\rm GeV}~~~p_{T3}>15~{\rm GeV}~~~p_{T4}>7~{\rm GeV},
\end{equation}
the invariant mass of the $e^+e^-$ and the $\mu^+\mu^-$ pairs should
be between
\begin{equation}
60~{\rm GeV}<M_{e^+e^-\!,\,\mu^+\mu^-}<105~{\rm GeV}
\end{equation}
and the invariant mass
of the $ZZ$ pair should fulfil
\begin{equation}
190~{\rm GeV}<M_{ZZ}<210~{\rm GeV}.
\end{equation}
With these cuts the NLL+LO result is 5.42 fb, which is about 2\% smaller
than the NLO one (5.51 fb).  The cross section from MC@NLO is about 11\%
larger (6.01 fb). This is mainly due to the fact that MC@NLO calculates the
cross section in the narrow width approximation, and therefore the cuts
on the invariant masses of the $e^+e^-$ and $\mu^+\mu^-$ pairs are
always fulfilled. As in the inclusive case,
the effect of scale variations on the rate is very small, of the order of $\pm 1\%$.

We point out that single-resonant contributions are neglected in our
calculation. We have used MadGraph/MadEvent \cite{madgraph} to check
that these contributions are indeed small and found that at LO the
effects are smaller than the permille level.  Effects from off-shell
photons $pp\to Z\gamma^*\to e^+e^-\mu^+\mu^-$ are larger. At NLO they
decrease the cross section by about 1\% with the cuts described above,
due to negative interference between the $Z$ boson and the photon.
The shapes of the distributions are, however, not significantly
changed. Hence, it is safe to neglect these two contributions in the
NLL+LO approximation with the cuts described above. For selection cuts
used in Higgs searches where its mass is smaller than the $ZZ$
threshold, the effects from off-shell photons cannot be neglected and
have to be included.

\begin{figure}[tb]
  \begin{center}
    \epsfig{file=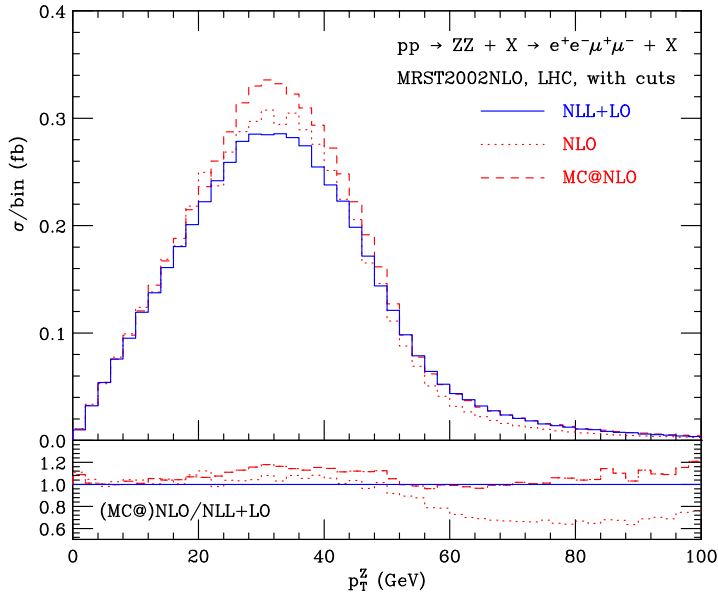, scale=0.60}
  \end{center}
  \vspace{-20pt}
  \caption{\label{fig:ptz_cuts}{\em Comparison of the transverse
      momentum spectra of one of the $Z$ at NLL+LO (solid) with NLO (dots)
      and MC@NLO (dashed) results. The lower part of the plot shows the NLO and MC@NLO
      results normalized to NLL+LO.}}
\end{figure}

In Fig.~\ref{fig:ptz_cuts} we show the $p_T$ distribution of one of
the $Z$ bosons, computed at NLL+LO, NLO and with MC@NLO.  Contrary to
the $\ptZZ$ spectrum, this distribution is well behaved at NLO but the
effect of resummation is still visible on its shape. This is evident
from the lower part of the plot, showing the NLO and MC@NLO result normalized to
NLL+LO.

\begin{figure}[tb]
  \begin{center}
    \epsfig{file=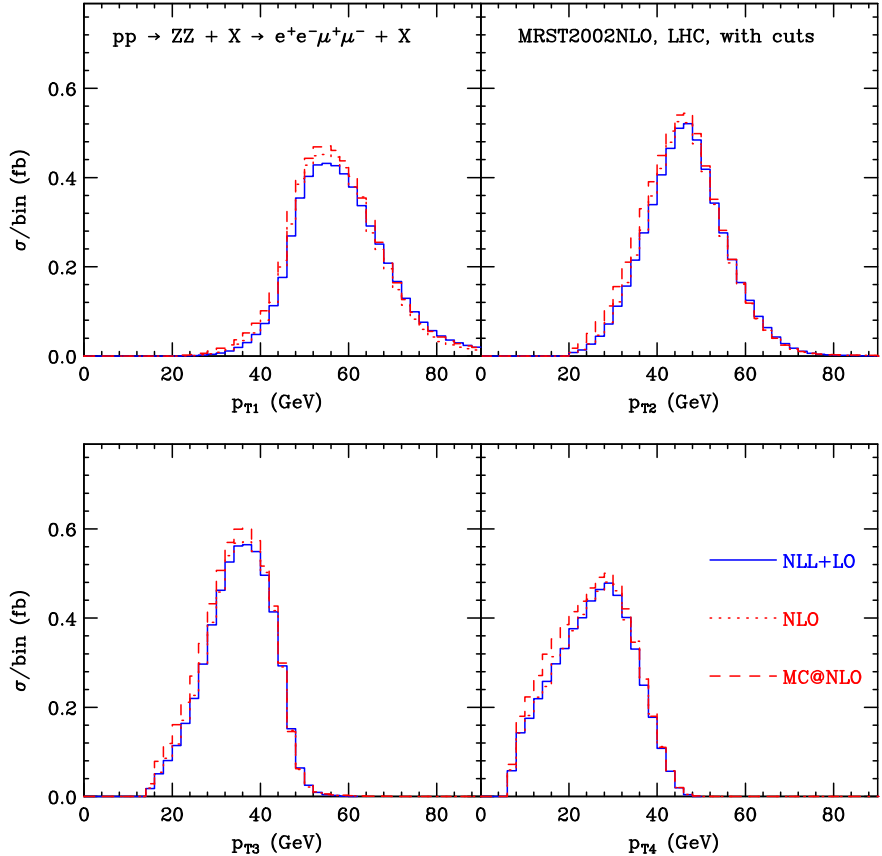, scale=0.80}
  \end{center}
  \vspace{-20pt}
  \caption{\label{fig:ptleptons_cuts}{\em Transverse momentum spectra of
      the leptons: NLL+LO (solid), MC@NLO (dashes) and NLO (dots).}}
\end{figure}

In Fig.~\ref{fig:ptleptons_cuts} we show the $p_T$ distributions of
the charged leptons, ordered according to decreasing $p_T$.  Here the
NLO prediction is in good agreement with the NLL+LO one. MC@NLO, however,
predicts slightly softer leptons.

The effect of scale variations is still very small for the above distributions. Only in the high-$p_T$
tail of the $p_T^Z$ distribution resummation scale variations give a visible effect,
being of about $\pm 10\%$ at $p_T^Z\sim M_Z$.

\begin{figure}[tb]
  \begin{center}
    \epsfig{file=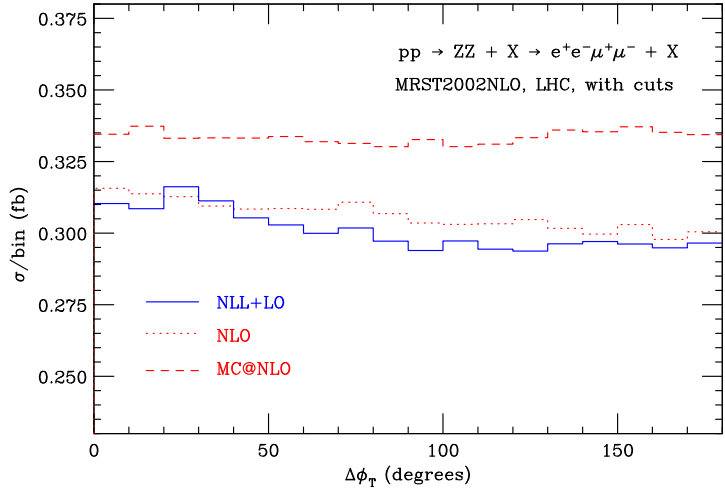, scale=0.60}
  \end{center}
  \vspace{-20pt}
  \caption{\label{fig:dphi_cuts}{\em Same as in
      Fig.~\ref{fig:ptleptons_cuts} but for the $\Delta\phi_T$
      distribution.}}
\end{figure}

In Fig.~\ref{fig:dphi_cuts} we consider the distribution in
the variable $\Delta\phi_T$ defined as follows. We consider the
separation between the $e^-$ and the $\mu^-$ where their momenta are
taken in the rest frame of their parent $Z$ boson, by neglecting all
the longitudinal components.
In this way the $\Delta\phi_T$ is manifestly
longitudinally invariant. As will be illustrated later, see
Sect.~\ref{sec:sb}, this angle is sensitive to the CP nature of a
Higgs boson resonance. Due to the fully transverse nature of this
angle, it can potentially be reconstructed also if only three leptons
are detected together with missing $E_T$.

We see that the shapes of the NLO and NLL+LO distributions are
qualitatively similar. Both decrease with increasing separation angle,
the NLL+LO prediction slightly more at small angles before it flattens
out, while the NLO prediction has a more constant slope. The
differences are, however, small. The effect of scale variations on the NLO and NLL+LO results
is again of the order or smaller than $1\%$.

We also plotted the prediction for
this angle by MC@NLO, although we remind the reader that MC@NLO does
not include spin correlations in the $Z$ decay.
Despite this fact, the shape of this distribution is not too different
from those obtained at NLO and NLL+LO.

\section{Signal and background}
\label{sec:sb}

In this Section we perform a consistent comparison of signal and
background cross sections for the Higgs search
in the $gg\to h\to ZZ\to e^+e^-\mu^+\mu^-$
channel at the LHC.  We consider a Higgs boson with mass $m_h=200$ GeV
and use the numerical program of
Refs.~\cite{Bozzi:2003jy,Bozzi:2005wk} to compute its transverse
momentum spectrum.  To be consistent with the background\footnote{In our simplified analysis, we consider only the $ZZ$
irreducible background.  We neglect other sources of reducible
background like $t{\bar t}$ and $Zb{\bar b}$ which are known to give a
much smaller contribution \cite{cmsnote}.}, we work at NLL+LO accuracy
and we generate a set of events containing a Higgs boson which is then
let decay using the MadGraph package \cite{madgraph}.  We use
the same cuts as in Sect.~\ref{sec:zz}.
 
\begin{figure}[tb]
  \begin{center}
    \epsfig{file=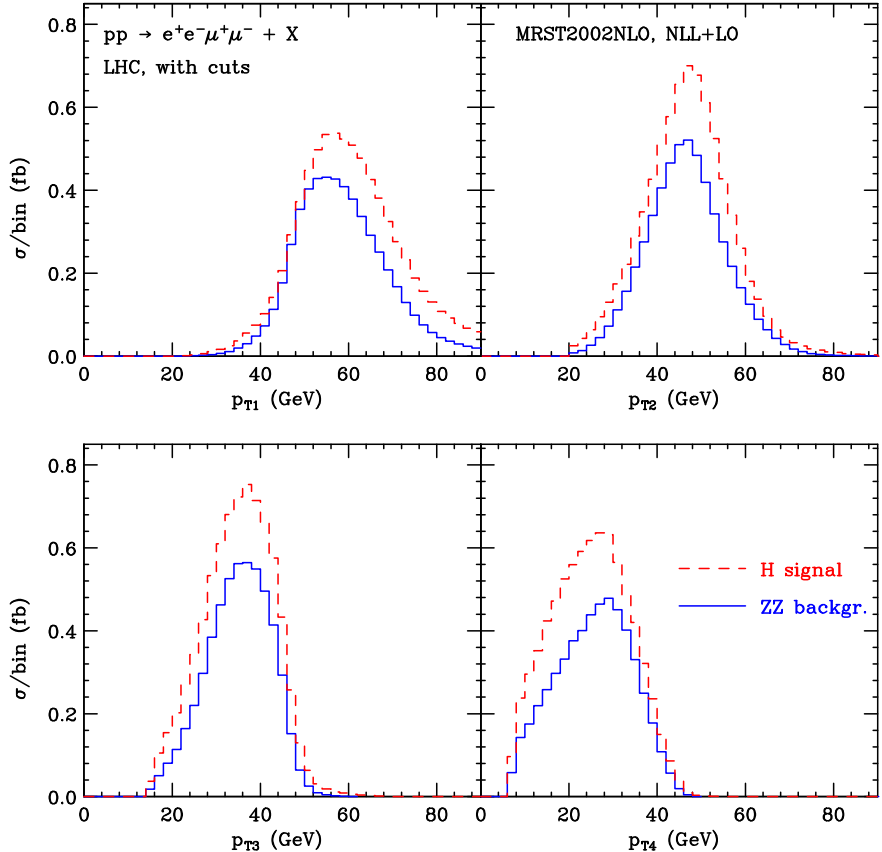, scale=0.80}
  \end{center}
  \vspace{-20pt}
\caption{\label{fig:ptleptonssb}{\em Leptons $p_T$ spectra for Higgs
signal (dashes) and $ZZ$ background (solid).}}
\end{figure}

The signal cross section is 7.74 fb.
Comparing with the
background we get $S/B=1.43$.
In Fig.~\ref{fig:ptleptonssb} we plot the $p_T$ spectra of the leptons
for signal and background. We see that the spectrum of the leading
lepton tends to be slightly harder for the signal, compared to the
background, whereas the opposite happens for the lepton with the
minimum $p_T$.

\begin{figure}[tb]
  \begin{center}
    \epsfig{file=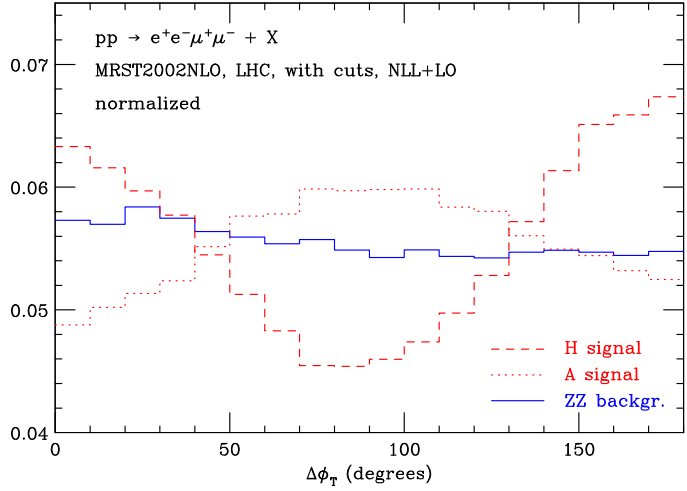, scale=0.60}
  \end{center}
  \vspace{-20pt}
\caption{\label{fig:dphisb}{\em The $\Delta\phi_T$ distribution for
scalar (dashes), pseudo-scalar (dots) and $ZZ$ background (solid) at NLL+LO.}}
\end{figure}

In Fig.~\ref{fig:dphisb} we plot the $\Delta\phi_T$ distribution
defined in Sect.~\ref{sec:zz} for signal and background.  Since this
distribution is expected to be sensitive to the CP nature of the Higgs, we also consider the case of a pseudo-scalar Higgs boson. 
As in the case of the scalar, the events are generated starting from
the transverse momentum spectrum at NLL+LO and then
letting the Higgs boson decay using
the MadGraph package \cite{madgraph}.
The computation of the spectrum for the pseudoscalar has been done
by using a modified version of the
numerical program of Refs.~\cite{Bozzi:2003jy,Bozzi:2005wk}, using
the results of Ref.~\cite{Kauffman:1993nv}\footnote{The spectrum
for the pseudoscalar at NLL+LO accuracy can be
easily obtained by using the fact that the real corrections
(in the large-$m_{top}$ approximation)
are the same as for the scalar. As such, the only difference from the case of the scalar is in the finite part of the virtual corrections \cite{Kauffman:1993nv}.}.

From Fig.~\ref{fig:dphisb} we see that the shape of
the distribution shows remarkable differences in the three cases.
As shown in Fig.~\ref{fig:dphi_cuts}, for the background the distribution is rather flat. On the contrary, for the pseudoscalar,
the distribution is peaked
at central values of $\Delta\phi_T$, whereas for the scalar the distribution has a minimum in this region.
We conclude that this angular variable
has a good discriminating potential to assess the CP nature of
the Higgs boson.

\begin{figure}[tb]
  \begin{center}
    \epsfig{file=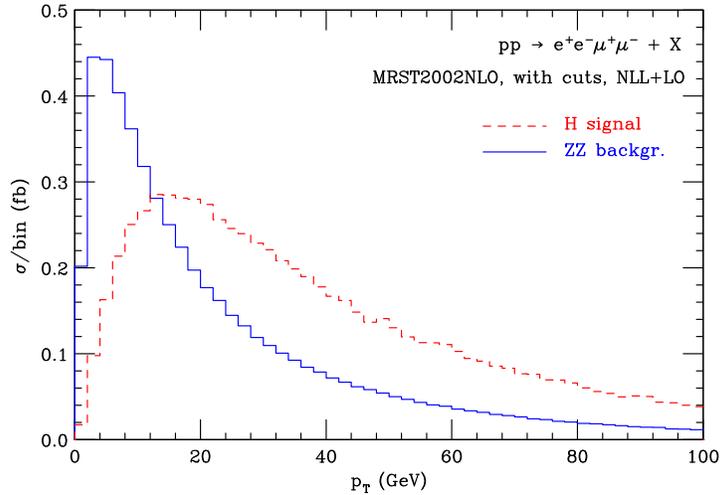, scale=0.60}
  \end{center}
  \vspace{-20pt}
\caption{\label{fig:ptzz_cuts}{\em Comparison of $p_T$ spectra of
signal and background at NLL+LO, when standard cuts are applied.}}
\end{figure}

We finally consider the possibility to apply an additional cut on the
total transverse momentum of the four leptons.  This idea is inspired
by a comparison of the transverse momentum spectra of the Higgs boson
and of the $ZZ$ pair in Fig.~\ref{fig:ptzz_cuts}.

We see from Fig.~\ref{fig:ptzz_cuts} that the Higgs signal is
definitely harder than the $ZZ$ background, being peaked at $p_T\sim
17$ GeV.  The $ZZ$ background is instead peaked at $p_T\sim 5$ GeV. As
such, a cut on the total transverse momentum of the leptons may
increase the statistical significance. Starting with the standard set
of cuts used in the rest of the paper, we compute the efficiency of
the additional cut by defining
\begin{equation}
\epsilon(\ptcut)=\sigma_{p_{T}>\ptcut}/\sigma\, .
\end{equation}

\begin{figure}[tb]
  \begin{center}
    \epsfig{file=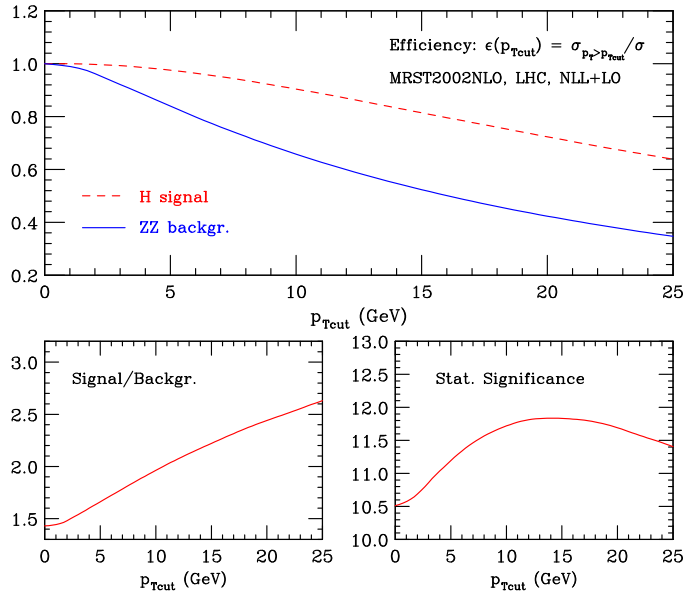, scale=0.6}
  \end{center}
  \vspace{-20pt}
\caption{\label{fig:eff}{\em The upper plot shows the NLL+LO efficiency as a function of
$\ptcut$ for the Higgs signal (dashed) and the
$ZZ$ background (solid). The lower left plot shows the
signal over background ratio and the lower right
the statistical significance for an
integrated luminosity of 10 fb$^{-1}$.}}
\end{figure}

In Fig.~\ref{fig:eff} we plot the efficiency as a function of
$\ptcut$ for the signal and the background.  We see that the
efficiency of this additional cut decreases more rapidly for the
background than for the signal.  As should be expected, the
resummation effect is crucial in this case.
The fixed order NLO efficiencies, not shown in Fig.~\ref{fig:eff},
become unphysically larger than unity for
small values of $\ptcut$.
Due to the fact that the efficiency of the background
decreases more rapidly
compared to the signal, the signal over background ratio increases
with increasing $\ptcut$, as can also be seen from the lower left
plot of Fig.~\ref{fig:eff}.  The lower right plot shows the
statistical significance
for an integrated luminosity of 10 fb$^{-1}$,
We observe that the statistical significance
is maximum when $\ptcut\sim 15$~GeV.
 
The latter point, however, requires a word of caution.
The predictions presented
in the present paper are based on resummed calculations obtained in a purely perturbative framework.
Intrinsic-$p_T$ effects are
known (see e.g. Ref.~\cite{Collins:va} 
and references therein)
to affect transverse-momentum distributions,
particularly
at small transverse momenta.
These effects
are not taken into account in our calculation.

As noted in Ref.~\cite{Bozzi:2005wk},
these non-perturbative effects
have the same qualitative impact
as the inclusion of higher-order logarithmic contributions, {\em i.e.}, they tend
to make the resummed $p_T$ distribution harder.
The quantitative results shown in Fig.~\ref{fig:eff} will certainly
depend on these effects, although the qualitative picture should not
change dramatically.

\section{Summary}
\label{summa}

Higgs boson production by gluon-gluon fusion, followed
by the decay mode $h\to ZZ\to 4$ leptons, provides the best
discovery channel at the LHC for Higgs masses above 180 GeV.  For a
precise determination of the properties of the Higgs resonance, such
as its mass and CP nature, detailed theoretical predictions for the
signal and backgrounds are necessary.

In this work we considered a Higgs boson with mass $m_h=200$ GeV and
performed the resummation of multiple soft-gluon emission for the $ZZ$ background.
We then compared the results with those for the signal
in the case of a (pseudo-)scalar Higgs boson. The effects from the
resummation of soft gluons are modest for observables like the
transverse momentum of the final state leptons. However,
the transverse momentum spectra of the $Z$ bosons and
of the $ZZ$ pair are sensitive to these effects.

An angle $\Delta\phi_T$ that is sensitive to the CP nature of the
Higgs signal is also introduced. This angle is defined in a fully
transverse way, such that it is longitudinally boost invariant. This
angle can potentially be reconstructed also if only three leptons
are detected together with missing $E_T$.

We also argued that an additional cut on the transverse
momentum of the $ZZ$ pair may significantly increase the signal over
background ratio and the statistical significance. The impact of resummation is of course crucial in this case. The above cut
could be helpful to claim an early discovery or to obtain
an easier determination of the nature of the discovered particle.

\subsection*{Acknowledgements}
We wish to thank Fabio Maltoni for valuable comments and
useful discussions and suggestions.  We would also like to thank Andrea Giammanco and Sasha Nikitenko for
enlightening discussions and Stefano Catani for comments on the manuscript. MG thanks the Center for Particle Physics and Phenomenology of Louvain University for the kind hospitality extended to him at various stages of this work. RF is partially supported by the Belgian
Federal Science Policy (IAP 6/11).

\end{document}